\documentclass{aa}

\usepackage{float}
\usepackage{graphicx}
\usepackage{longtable}
\usepackage{threeparttable}
\usepackage{natbib}
\usepackage{txfonts}
\usepackage{color}
\usepackage{enumerate}
\usepackage{listings}
\usepackage{bm}
\usepackage{hyperref}
\usepackage{comment}

\defcitealias{2020A&A...641A.127R}{R20}
\defcitealias{2023A&A...669A.125T}{T23}
\usepackage{enumitem}
\setlist[enumerate]{label=\textbullet}

\hypersetup{
  colorlinks=true,
  pdfborder=2 2 0.,
  linkcolor=blue,
  anchorcolor=black,
  citecolor=blue,
  urlcolor=black,
}

\begin{document} 

\title{APOGEE chemical abundances of stars in the MW satellites Fornax, Sextans, Draco and Carina}

\titlerunning{MW satellites}
\authorrunning{Xu et~al.}

\author{Cheng Xu\inst{1,2}, Yi Qiao\inst{1,2}, Baitian Tang\inst{1,2}, Jos\'e G. Fern\'andez-Trincado\inst{3,4}, Zhiqiang Yan \inst{5,6}, Ruoyun Huang\inst{1,2}, and Doug Geisler \inst{7,8}}

\institute{Department of Astronomy, School of Physics and Astronomy, Sun Yat-sen University, Zhuhai, Guangdong Province, China \\
            \email{tangbt@sysu.edu.cn}
\and{CSST Science Center for the Guangdong-Hong Kong-Macau Greater Bay Area, Zhuhai 519082, China}
\and{Universidad Cat\'olica del Norte, N\'ucleo UCN en Arqueolog\'ia Gal\'actica - Inst. de Astronom\'ia, Av. Angamos 0610, Antofagasta, Chile}
\and{Universidad Cat\'olica del Norte, Departamento de Ingenier\'ia de Sistemas y Computaci\'on, Av. Angamos 0610, Antofagasta, Chile}
\and{School of Astronomy and Space Science, Nanjing University, Nanjing 210093, PR China}
\and{Key Laboratory of Modern Astronomy and Astrophysics (Nanjing University), Ministry of Education, Nanjing 210093, PR China}
\and{Departamento de Astronom\'{i}a, Casilla 160-C, Universidad de Concepci\'{o}n, Concepci\'{o}n, Chile}
\and{Departamento de Astronom\'ia, Facultad de Ciencias, Universidad de La Serena. Av. Ra\'ul Bitr\'an 1305, La Serena, Chile}
}

\abstract

\abstract{\hspace{0.2cm}During its evolution, the Milky Way (MW) incorporated numerous dwarf galaxies, particularly low-mass systems. The surviving dwarf galaxies orbiting the MW serve as exceptional laboratories for studying the unique properties of these systems. Their metal-poor environments and shallow gravitational potentials likely drive significant differences in star formation and star cluster properties compared to those in the MW. Using high-quality near-infrared spectra from the APOGEE survey, we determined abundances of Fe, C, N, O, Mg, Al, Si, Ca, Ti, Cr, Mn, Ni, and Ce for 74 stars in four MW satellite dwarf galaxies: Fornax, Sextans, Draco, and Carina. 
Our analysis reveals that the distribution of $\alpha$ elements (e.g., [Si/Fe]) strongly correlates with galaxy luminosity (and hence mass), underscoring the critical role of galaxy mass in shaping chemical evolution. These dwarf galaxies exhibit [Al/Fe$]\sim -0.5$, which is comparable to those of the metal-poor stars in the MW. Additionally, we identified nitrogen-rich field stars in the Fornax dwarf galaxy, which display distinct metallicities compared to its known globular clusters (GCs). If these stars originated in GCs and subsequently escaped, their presence suggests we are observing relics of destroyed GCs, offering possible evidence of cluster disruption.
}


\maketitle
%

\section{Introduction}

\hspace{0.3cm}Galaxies represent the fundamental building blocks of the universe.  Their observed scaling relations, e.g., the mass-luminosity relation \citep[e.g.,][]{2003ApJS..149..289B}, and the mass-metallicity relation \citep[e.g.,][]{1979A&A....80..155L, 2012A&A...547A..79F}, play a pivotal role in deciphering galaxy formation and evolution across diverse cosmic environments characterized by variations in age, metallicity, and redshift. Nevertheless, the underlying physical mechanisms driving these processes remain subjects of ongoing scientific debate. In this context, the ability to resolve galaxies into their constituent stars provides the most direct observational approach for probing their assembly histories.

\hspace{0.2cm}The hierarchical galaxy formation paradigm suggests that dwarf galaxies originate within dark matter subhalos, subsequently merging to assemble larger galactic systems like the Milky Way (MW). This framework is observationally supported by the numerous dwarf satellites identified in the MW’s vicinity \citep{1971ARA&A...9...35H,1989ARA&A..27..139H, 2003AJ....125.1926G, 2009ARA&A..47..371T, 2012AJ....144....4M,2013ApJ...765L..15I, 2014ApJ...789..147W,2019ARA&A..57..375S}. Recent discoveries have further revealed both tidally disrupting systems, such as the Sagittarius (Sgr) stream \citep[][]{Majewski2003,2010ApJ...718.1128L}, and fully disrupted remnants including the Gaia-Sausage-Enceladus \citep[GSE;][]{2018MNRAS.478..611B,2018Natur.563...85H}, and the Sequoia \citep[e.g.,][]{2019MNRAS.488.1235M} substructures.
These systems display remarkable diversity in mass distribution, morphological configuration, and stellar population properties. Through comparative analysis with numerical simulations, they offer critical insights into halo assembly processes, the environmental regulation of galaxy formation, and their dynamical interactions throughout cosmic history.

\hspace{0.2cm}The stellar chemical abundance patterns of galaxies serve as a fossil record of their evolutionary history. For example, stars in dwarf galaxies show systematically lower metallities than those of MW stars. Furthermore, the characteristic ``knee'' in the [$\alpha$/Fe]-[Fe/H] trend provides a diagnostic tool for estimating the star formation history (hereafter SFH) in dwarf galaxies \citep[e.g.,][]{1979ApJ...229.1046T,2003AJ....125..684S,2006ApJ...653.1145K,2014ApJ...796...38N,2020AJ....159...46K,2020ApJ...895...88N}. Building on this framework, \citet{2021ApJ...923..172H} proposed the potential occurrence of a second starburst episode in several nearby MW satellite galaxies, inferred from chemical evolution modeling of their [$\alpha$/Fe]–[Fe/H] distributions. 
High-precision abundance measurements also enable investigations into variations in the stellar initial mass function (IMF) across galactic environments \citep[e.g.,][]{2013ApJ...778..149M,2017ApJ...845..162H,2018ApJ...859L..10C,Yan2020}. These chemical signatures collectively offer critical insights into the diverse physical processes governing galaxy assembly and evolution.

\hspace{0.2cm}Clustered star formation is ubiquitously observed across cosmic time, manifesting in both the low-redshift (low-z) Universe \citep{2024arXiv241107424P} and the high-z Universe \citep[e.g.,][]{2024Natur.632..513A}. Surviving star clusters serve as crucial tracers of galactic star formation histories. These stellar systems are typically classified into open clusters and globular clusters (GCs), with the latter exhibiting distinctive chemical abundance patterns characterized by N, Na, and Al enhancements alongside C, O, and Mg depletion \citep{2010A&A...516A..55C,2015ApJ...808...51M,2015AJ....149..153M,2018ARA&A..56...83B,2017MNRAS.465...19T,2018ApJ...855...38T,2021ApJ...908..220T,2021ApJ...906..133L} --- a phenomenon recognized as multiple populations (MPs). Our ongoing research initiative, ``Scrutinizing {\bf GA}laxy-{\bf ST}a{\bf R} cluster coevoluti{\bf ON} with chem{\bf O}dyna{\bf MI}cs ({\bf GASTRONOMI})'', employs multi-wavelength photometric and spectroscopic data to unravel the coevolutionary relationships between the MW, its satellite dwarf galaxies, and their stellar clusters. 

Extending this investigation to the Sculptor dwarf galaxy (Scl), \citet[][hereafter T23]{2023A&A...669A.125T} reported no detectable GCs or chemically peculiar stars indicative of GC dissolution, and the modeling analysis of APOGEE chemical abundance data yields a top-light IMF for Scl characterized by a high-mass slope of $-2.7$. This paucity of massive stars aligns with the observed dearth of compact star clusters, which typically serve as nurseries for high-mass star formation.

\hspace{0.2cm}As GCs orbit their host galaxies, they gradually lose stars to the field population through dynamical processes, with some clusters eventually dissolving entirely.  Identifying these GC escapees is important for modeling galaxy-star cluster interactions, yet their kinematic resemblance to halo stars makes them challenging to detect. Recent studies have uncovered a substantial population of nitrogen-rich (N-rich) field stars in the MW \citep[e.g.,][]{2015A&A...575L..12L,2016ApJ...825..146M,2017MNRAS.465..501S,2016ApJ...833..132F,2017ApJ...846L...2F,2019MNRAS.488.2864F,2019ApJ...871...58T,
2021ApJ...913...23Y,2022A&A...663A.126F}, which shared the most significant feature of MPs in GCs $-$ N-enhancement. These stars are widely regarded as the strongest candidates for escaped GC members. Moreover, N-rich field stars have been detected  in the Magellanic clouds (MCs) and Sgr \citep{2020ApJ...903L..17F, 2021A&A...648A..70F}, extending the evidence for GC dissolution to diverse galactic environments. By combining observations of extant GCs with their dispersed stellar remnants, we can construct a more complete picture of GC evolution, dissolution histories, and their role in shaping galactic stellar populations.

\hspace{0.2cm}How do lower-mass Milky Way satellites contribute to our understanding of galactic chemical evolution? Can N-rich field stars be identified in these systems, and what do they reveal about the interplay between star cluster dynamics and galaxy evolution? In this study, we present detailed chemical abundance measurements for up to 13 elements across four dwarf spheroidal galaxies: Fornax (Fnx), Draco (Dra), Carina (Car), and Sextans (Sex).  Section~\ref{sec:Data} outlines the data acquisition and reduction. In Section~\ref{sec: result}, we systematically compare abundance patterns for $\alpha$-elements (O, Mg, Si), iron-peak species (Cr, Mn, Ni), and neutron-capture elements (e.g., Ce), highlighting both commonalities and divergences across these satellite systems. We further examine radial abundance gradients and investigate the presence of N-rich field stars, discussing their potential origins and implications in Section~\ref{sec:discussion}. Finally, our conclusions and a summary of key findings are outlined in Section~\ref{sec: summary}.

\section{Data reduction}
\label{sec:Data}

\subsection{Observational data}
\label{sec:Obv}

\hspace{0.3cm}The near-infrared (H-band, $\lambda$ = 1.51 -- 1.69 $\mu$m), high-resolution (R $\sim$ 22,500) spectra are provided by the Apache Point Observatory Galactic Evolution Experiment (APOGEE, \citealt{2017AJ....154...94M}), which was part of the Sloan Digital Sky Survey III (SDSS-III, \citealt{2011AJ....142...72E}) and SDSS-IV \citep{2017AJ....154...28B}. 
The APOGEE survey were carried out with two telescopes: the 2.5 m telescope at the Apache Point Observatory \citep{2017AJ....154...28B} and the 2.5 m du Pont telescope at the Las Campanas Observatory \citep{1973ApOpt..12.1430B}. The APOGEE survey sample primarily consists of red giant stars (RGB) across the MW \citep{2017AJ....154...94M,2021AJ....162..303S,2021AJ....162..302B}. To date, APOGEE has observed nearly a million stars, covering the MW and neighboring galaxies. Furthermore, APOGEE data reduction software is used to simplify multiple 3D raw data cubes into well-sampled, calibrated combined 1D spectra \citep{2015AJ....150..173N}. Stellar parameters are derived using the FERRE code \citep{2006ApJ...636..804A}, which finds the best solution by comparing the observed spectra with the libraries of theoretical spectra \citep{2015AJ....149..181Z}.

\begin{figure*}[htbp]
    \sidecaption
    \includegraphics[width=12cm]{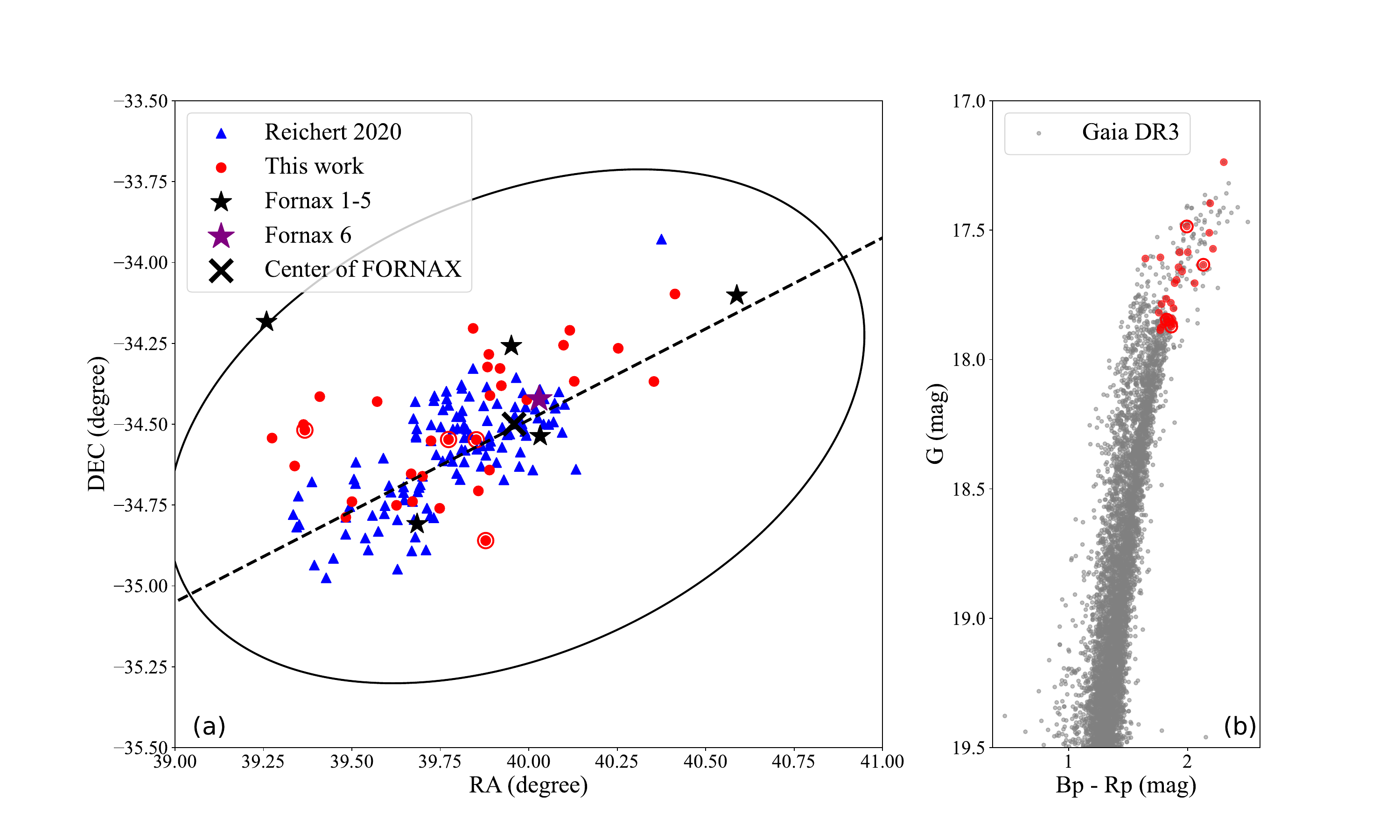}
    \caption{Target information. (a): Spatial distribution of target stars in Fnx dwarf galaxy. The black ellipse indicates the tidal radius of Fnx and its center is labelled with black cross. Target stars from \citeauthor{2020A&A...641A.127R} (\citeyear{2020A&A...641A.127R}, hereafter \citetalias{2020A&A...641A.127R}) and this work are marked as blue triangles and red dots, respectively.  The N-rich stars are further marked with red circles. The black stars represent GC Fnx 1-5 while the purple star represents GC Fnx 6.  (b): Color--magnitude diagram of Fnx stars based on $Gaia\ DR3$ (grey dots).}
    \label{fig:FornaxCMD}%
\end{figure*}

\hspace{0.2cm}With 17th data releases of SDSS (DR 17, \citealt{2022ApJS..259...35A}) to date, dwarf galaxies in the vicinity of the MW have been observed multiple times, and the signal-to-noise ratios per pixel (S/N) have reached a level sufficient for chemical analysis. In this work, we cross-matched the catalogs of \citet{2018A&A...616A..12G} and APOGEE DR17 to select potential member stars of Fnx, Dra, Car, and Sex dwarf galaxies. To reduce the uncertainties of derived chemical abundances, we excluded spectra with $S/N< 70$. Furthermore, we examined the radial velocities (RVs) of these candidates (Table \ref{Table:  Full Dwarf members stellar parameters}), finding that the RVs of all member stars fall within $3\sigma$ range. Therefore, the reliability of these member stars is confirmed by their spatial positions, proper motions, and RVs. Their positions in the color–magnitude diagram are also checked. There are 32 members in Fnx, 14 members in Dra, 19 members in Car, and 8 members in Sex. The locations of our Fnx sample stars and their positions in the color–magnitude diagram are plotted in Fig. \ref{fig:FornaxCMD}.

\subsection{Stellar parameters and chemical abundances}

\hspace{0.3cm}In this work, we determined the photometric effective temperature ($T_{\rm eff}$) using  2MASS photometry, following the method outlined by \citet{2009A&A...497..497G}. The final $T_{\rm eff}$ values were derived from the dereddened color indices ($J-K$), using the extinction map from \citet{1998ApJ...500..525S}. The surface gravity (log $g$) values were estimated from the canonical equation: 
\begin{equation}
    \log(\frac{g}{g_{\odot}})=\log(\frac{M}{M_{\odot}})+4\log(\frac{T}{T_{\odot}})-\log(\frac{L}{L_{\odot}})
\end{equation}
where the stellar mass is assumed to be 0.8 solar mass. To estimate the bolometric luminosities, we apply the bolometric correction procedure described by \citet{1999A&AS..140..261A}. The derived values of $T_{\rm eff}$ and $\log g$ are presented in Table B1. The uncertainties of $T_{\rm eff}$ are dominated by those of the $T_{\rm eff}$-($J-K$) relation, which is 94 K. The $T_{\rm eff}$ uncertainties are propagated to the estimation of $\log g$ uncertainties. 

\hspace{0.2cm}Chemical abundances under the assumption of local thermal equilibrium (LTE), were analyzed using the Brussels Automatic Stellar Parameter (BACCHUS) code \citep{2016ascl.soft05004M}. Built upon the radiative transfer code, Turbospectrum \citep{1998A&A...330.1109A, 2012ascl.soft05004P}, BACCHUS interpolates model atmospheres from the MARCS model grids \citep{2008A&A...486..951G}. The code's meticulous line selection and inspection capabilities were essential for analyzing the weak spectral lines characteristic of the metal-poor stars in our sample.
For each absorption line, the abundance derivation began with a sigma-clipping selection of continuum regions, followed by a linear fit. This step identifies and rejects poor fits caused by features like bad pixels or abrupt spectral drops. The abundance is then estimated by comparing the observed and synthetic spectra using four independent methods: (1) $\chi^2$ minimization of the overall fit; (2) comparison of the line core intensity; (3) equivalent width analysis; and (4) spectral synthesis (see Fig. \ref{fig:lines} for example spectra). Each method is assigned a quality flag. To ensure robustness, we only accepted an abundance measurement if all four methods yielded a satisfactory fit; otherwise, the line was rejected.  Following the recommendation of \cite{2016ascl.soft05004M}, we adopted the abundance derived from the $\chi^2$ method as the final value for each line available. This choice is motivated by its overall robustness in matching the full spectral line profile.

\hspace{0.2cm}Given the prevalence of CO, CN, and OH molecular lines in the APOGEE spectral range, we determined the abundances of C, N, O, and Fe in a self-consistent, iterative procedure to account for their inter-dependencies, following \citet{2013ApJ...765...16S}. We first derived the oxygen abundance from $^{16}$OH lines, then carbon from $^{12}$C$^{16}$O, and nitrogen from $^{12}$C$^{14}$N, before determining the iron abundance from Fe I lines.
With these key elements established, we performed a line-by-line analysis of atomic lines for Mg, Al, Si, Ca, Ti, Cr, Mn, Ni, and Ce. While strong lines for Mg, Al, and Si allowed for a comprehensive analysis, many lines for Cr, Mn, Ni, and Ce, as well as some molecular $^{12}$C$^{16}$O and $^{12}$C$^{14}$N lines, were too weak for reliable measurement. These were carefully inspected and excluded from the final abundance set. The resulting chemical abundances are presented in Table B2, with solar abundances adopted from \citet{2005ASPC..336...25A} as our reference. Other elements, such as sodium (Na), were not analyzed due to their weak and unreliable absorption lines in our data.

\hspace{0.2cm}We first estimated the uncertainties of chemical abundances caused by individual stellar parameter error:  $\sigma_{\Delta \mathrm{T_{\rm eff}}}, \sigma_{\Delta \log \rm{g}}$, and $\sigma_{\Delta {\rm [Fe/H]}}$, where their typical errors were set as: $\Delta \mathrm{{T}_{\rm {eff}}} { = 100}\,{\rm K}$, $\Delta \log \rm{g} = 0.1\,{\rm dex}$, $\Delta \mathrm{[Fe/H]} = 0.05\,{\rm dex}$. 
Subsequently, the total estimated error is : $\sigma_{\rm tot} = [ (\sigma_{\Delta \mathrm{T_{\rm eff}}})^2 + (\sigma_{\Delta \log \rm{g}})^2 + (\sigma_{\Delta {\rm [Fe/H]}})^2 ]^{1/2}$. The errors for Star \#10 are shown in Table \ref{Table: abu_err} as an example. 

\begin{table}

\caption{Errors of chemical abundances propagated from atmospheric parameters for star \#10. }

\label{Table: abu_err}
\centering                                      
\setlength{\tabcolsep}{5pt}
\begin{tabular}{c c c c c}
\hline
\hline\\[-3pt]
        Element  & $\sigma_{\Delta T_{\rm eff}}$ & $\sigma_{\Delta \log g}$  & $\sigma_{\Delta {\rm [Fe/H]}}$  & $\sigma_{\rm tot}$ \\
\hline\\[-4pt]
{\rm [C/Fe]} &  0.02  &  0.03  & 0.02  & 0.04  \\ 
{\rm [N/Fe]} &  0.13  &  0.01  & 0.06  & 0.14  \\ 
{\rm [O/Fe]} &  0.18  &  0.01  & 0.04  & 0.19  \\ 
{\rm [Mg/Fe]} &  0.04  &  0.04  & 0.01  & 0.06  \\ 
{\rm [Al/Fe]} &  0.04  &  0.02  & 0.01  & 0.05  \\ 
{\rm [Si/Fe]} &  0.01  &  0.01  & 0.02  & 0.03  \\ 
{\rm [Ca/Fe]} &  0.05  &  0.01  & 0.01  & 0.05  \\ 
{\rm [Ti/Fe]} &  0.13  &  0.01  & 0.02  & 0.13  \\ 
{\rm [Cr/Fe]} &  0.15  &  0.01  & 0.04  & 0.16  \\ 
{\rm [Mn/Fe]} &  0.04  &  0.01  & 0.01  & 0.04  \\ 
{\rm [Ni/Fe]} &  0.01  &  0.03  & 0.02  & 0.04  \\ 
{\rm [Ce/Fe]} &  0.03  &  0.03  & 0.03  & 0.05  \\ 

\hline
\end{tabular}

\end{table}

\begin{figure*}[htbp]
    \sidecaption
    \includegraphics[width=12cm]{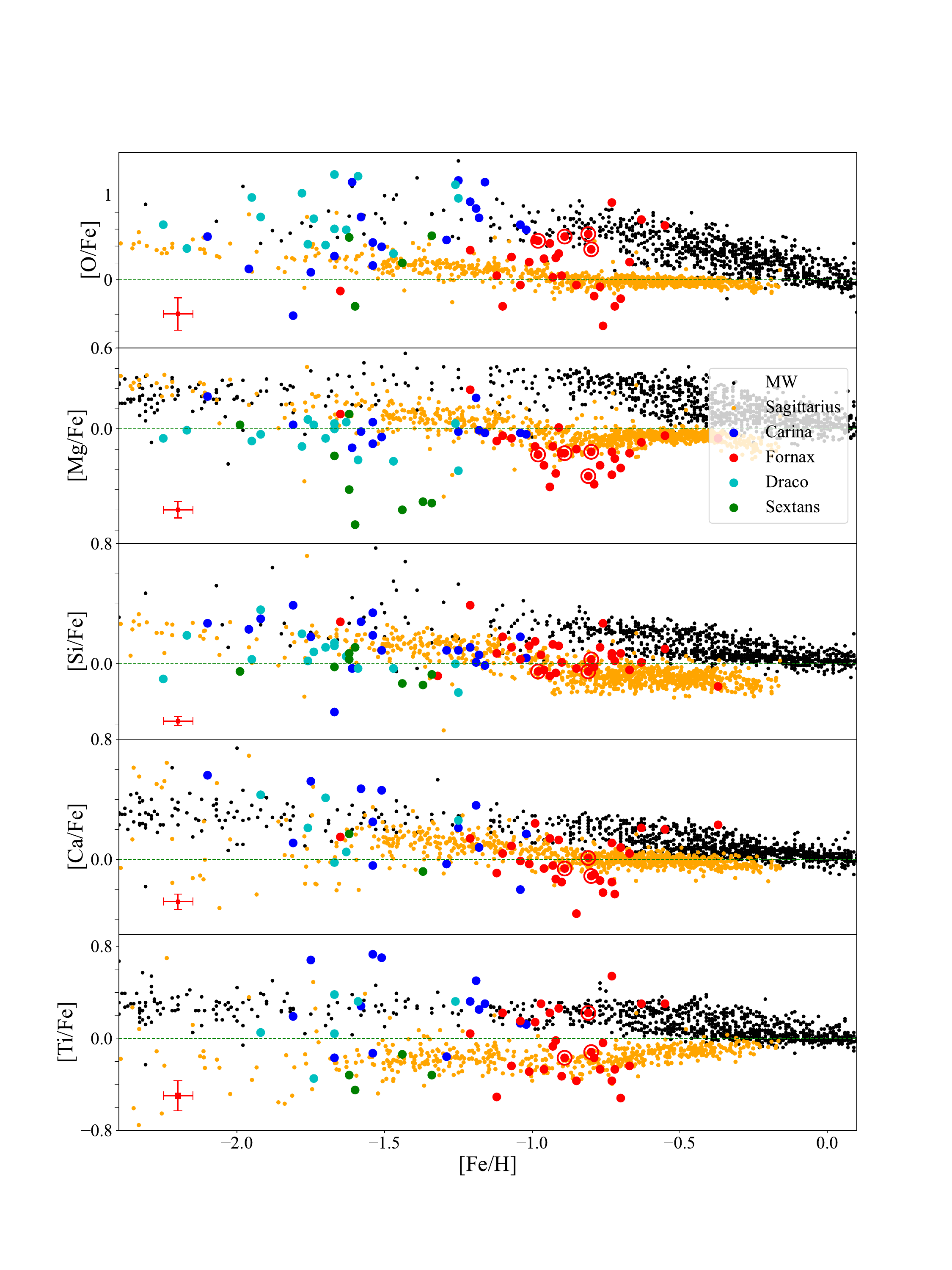} 
    \caption{[$\alpha$/Fe] vs. [Fe/H]. Fnx, Dra, Car and Sex stars from this work are labeled as red, cyan, blue and green stars respectively. The error bar in each panel indicates the typical error of available measurements. Black dots correspond to MW stars from the halo \citep{2000AJ....120.1841F,2004A&A...416.1117C,2005A&A...439..129B,2013ApJ...762...27Y,2014AJ....147..136R}, and MW stars from the disc \citep{2003MNRAS.340..304R, 2006MNRAS.367.1329R, 2014A&A...562A..71B}. Orange dots represent Sgr stars from \citet{2021ApJ...923..172H}. Purple squares represent GCs in Fnx from \citet{2022A&A...660A..88L}}
    \label{fig: alpha}%
\end{figure*}

\begin{figure}[htbp]
    \raggedright
    \includegraphics[width=9cm,height=6.2cm]{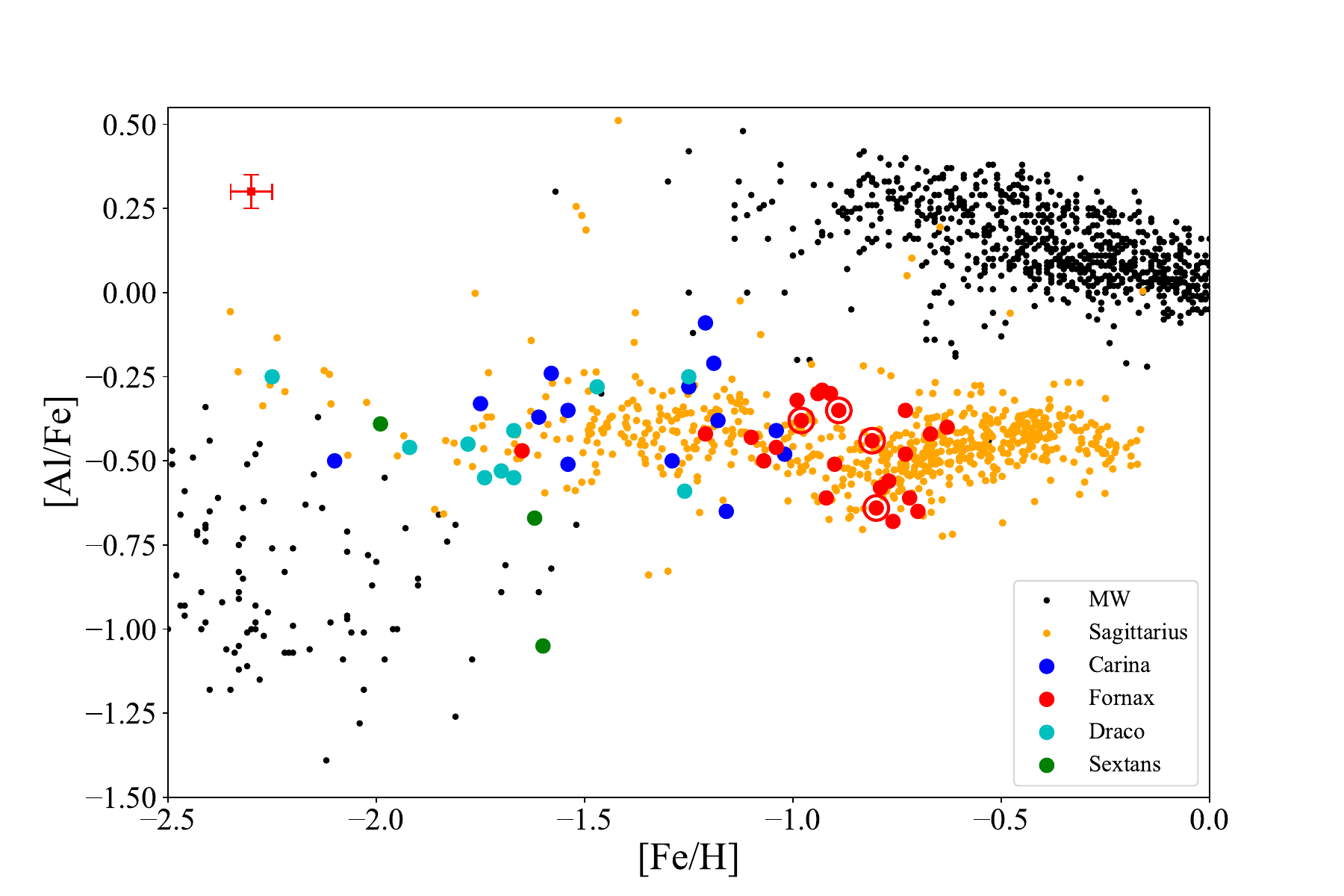} 
    \caption{[Al/Fe] vs. [Fe/H] relations. Symbols are the same as in Fig.~\ref{fig: alpha}.  }
    \label{fig: Al-Fe}
\end{figure}

\section{Results}
\label{sec: result}

\subsection{$\alpha$--elements}
\label{sec: result-alpha}

\hspace{0.3cm}According to nucleosynthetic models \citep{2020ApJ...900..179K}, $\alpha$-elements are primarily synthesized in exploding massive stars (Type II supernovae, SNe II), while iron is produced by both Type Ia supernovae (SNe Ia) and SNe II. Since the timescale of SNe II ($\sim$ Myr) is much shorter than that of SNe Ia ($\sim 100$ Myr), the interstellar medium (ISM) is contaminated by $\alpha$-rich materials at earliest time. Therefore, high $\alpha$ abundances are generally observed in low metallicity stars. After the onset of SNe Ia, so-called `SNe Ia delay time', the ejection of large amounts of iron continues to increase the [Fe/H], but decreases the [Mg/Fe] of the ISM and thus the later formed stars. This point of decline, termed the ``knee'' in the $[\alpha/{\rm Fe}]$ versus ${\rm [Fe/H]}$ trend, is influenced by the galaxy's star formation rate (SFR) and mass: generally the $\alpha$ knee appears at a lower metallicity in lower mass galaxies since they have a lower SFR \citep{2009ARA&A..47..371T}. 

The relationship between $[\alpha/{\rm Fe}]$ and ${\rm [Fe/H]}$ is illustrated in Fig. \ref{fig: alpha}, where background data include stars from the Galactic halo, disk, and dwarf galaxies. For each galaxy, we generally see a relatively constant $[\alpha/Fe]$ at low metallicity, which begins to decline above a certain metallicity, commonly referred as the $\alpha$ knee. However, the data are more complex than expected due to the uncertainties and systematics.

The Ti abundance measurements show larger dispersion due to the relatively weak absorption lines. Meanwhile, the O abundances also present a fairly large scattered distribution, particularly for metal-poor stars, which could be attributed to their iterated derivation procedures between C, N, O related molecular lines needed to achieve molecular equilibrium \citepalias{2023A&A...669A.125T}. Note that a few metal-rich Fnx stars ([Fe/H$]>-0.75$) show relatively high [O/Fe], probably due to their AGB nature with enhanced [Ce/Fe] (See Section \ref{sec: result-Neu}).
Our measured Mg abundances are systematically lower ($\sim 0.2$ dex) than those derived from the optical spectra, as we pointed out before. 

Interestingly, a large portion of stars in Sex show low [Mg/Fe] ($\sim -0.6$), which may be related to the second knee implied by \citetalias{2020A&A...641A.127R}  and \cite{theler_2020A&A...642A.176T}. We will further discuss the correlation between galaxy mass and $\alpha$-element distribution in Section \ref{sec:alphaknee}.

\subsection{Odd-Z element: Al}
\label{sec: result-Al}

\hspace{0.3cm}Al is primarily synthesized in massive stars \citep{1971ApJ...166..153A, 1995ApJS..101..181W} and is released into the ISM via SNe II. Its yield may depend on the metallicity of the progenitor stars \citep[e.g.,][]{2019ApJ...874..102W}. Al also can be generated via high temperature hydrogen burning through the Mg-Al cycle \citep{arnould1999}, which can be found in stellar objects like Wolf-Rayet stellar winds \citep{2006ApJ...647..483L} and AGB stars.

As shown in Fig. \ref{fig: Al-Fe}, the Al abundances of our studied dwarf galaxies are systematically lower than those from the MW disk ([Fe/H$]>-1.5$). This can also be found between MW in-situ and accreted globular clusters \citep{Lin2025} and field halo stars \citep{2025A&A...703A.256E}. However, they are comparable to some MW halo stars ([Fe/H$]<-1.5$). It is speculated that the early evolution of [Al/Fe] is similar between a MW-like galaxy and a GSE-like galaxy \citep{Horta2021}.  
No clear trend is seen in Car, Dra, and Sex, given that the sample size of each dwarf galaxy is small. We do find a vague trend that [Al/Fe] of Fnx stars decreases with increasing metallicity in the more metal-poor regime ([Fe/H$]<-0.75$), but start flattening in the more metal-rich regime. This possible flattening feature is also vaguely found in [Mg/Fe] and [Si/Fe]. \citet{2021ApJ...923..172H} suggests such a feature is caused by a secondary burst of star formation.

\begin{figure}[htbp]
    \raggedright
    \includegraphics[width=9.1cm,height=11.1cm]{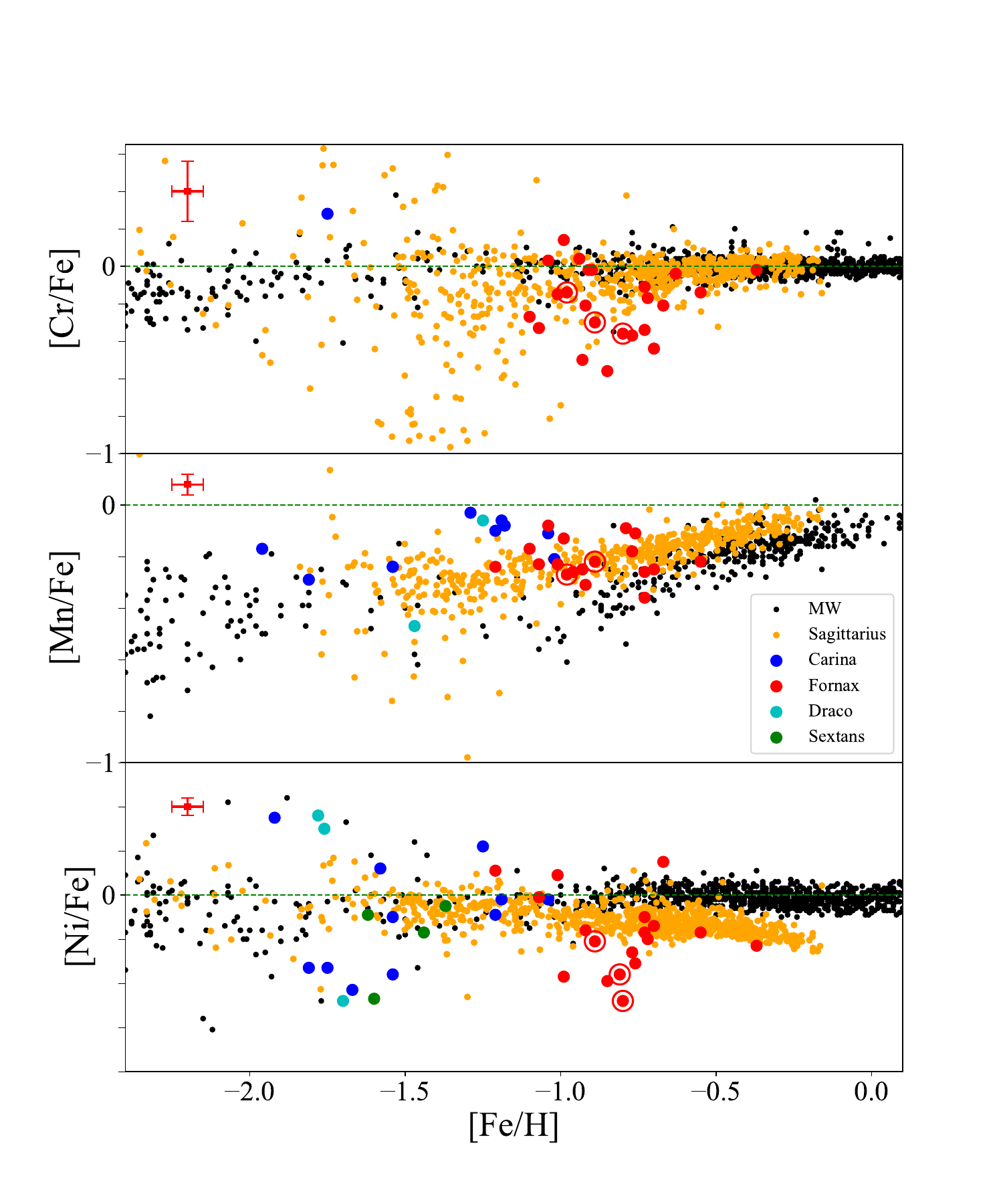} 
    \caption{Abundances of iron peak elements vs. [Fe/H].  Symbols are the same as in Fig. \ref{fig: alpha}.}
    \label{fig: Iron-Fe}
\end{figure}

\subsection{Iron-peak elements}
\label{sec: result-Iron}

\hspace{0.3cm}Analyses of solar elemental abundances reveal that iron-peak elements are actually contributed by both SNe Ia and core-collapse SNe II \citep{1995ApJS..101..181W}. Due to differences in star formation timescales and histories between the MW and dwarf galaxies, the iron-peak element abundances also vary accordingly.

The atomic lines for Cr, Mn, and Ni are relatively weak in the APOGEE spectra, making it impossible to obtain abundances for all sample stars. 
[Cr/Fe] generally shows a flat trend as a function of metallicity for MW, Sgr, and Fnx (Fig. \ref{fig: Iron-Fe}). The Sgr and Fnx stars show greater dispersion in Cr, especially for more metal-poor stars, probably due to their weaker lines and lower SNR in the NIR spectra. 

For MW and Sgr, [Mn/Fe] are almost constant ($\sim -0.3$) in the range of [Fe/H$]<-1.0$, and increase in the more metal-rich regime. Based on the limited measurements, Car stars show a similar flat trend ($\sim -0.2$) as a function of metallicity for [Fe/H$]<-1.0$; Fnx stars show similar [Mn/Fe] values as Sgr stars between $-1.0<[$Fe/H$]<-0.6$. However, it is difficult to tell its trend with metallicity given the limited metallicity range and large scatter.

No obvious [Ni/Fe] trend is found in the metal-poor regime ([Fe/H$]<-1.3$) for any aforementioned galaxies, due to their large abundance scatter. However, in the metal-rich regime ([Fe/H$]>-1.3$), Sgr stars show lower [Ni/Fe] compared to MW stars, and Fnx stars seem to show even lower values. In this regime, [Ni/Fe] decreases as the metallicity increases. Similar results are found in the optical counterpart \citepalias{2020A&A...641A.127R}. Theoretical studies suggest a significant contribution of sub-Chandrasekhar mass channels to SNe Ia in Fnx, possibly with additional progenitor mass variation related to SFH \citep{Sanders2021,Nissen2024}.

\begin{figure}
    \raggedright
    \includegraphics[width=9.0cm,height = 5.9cm]{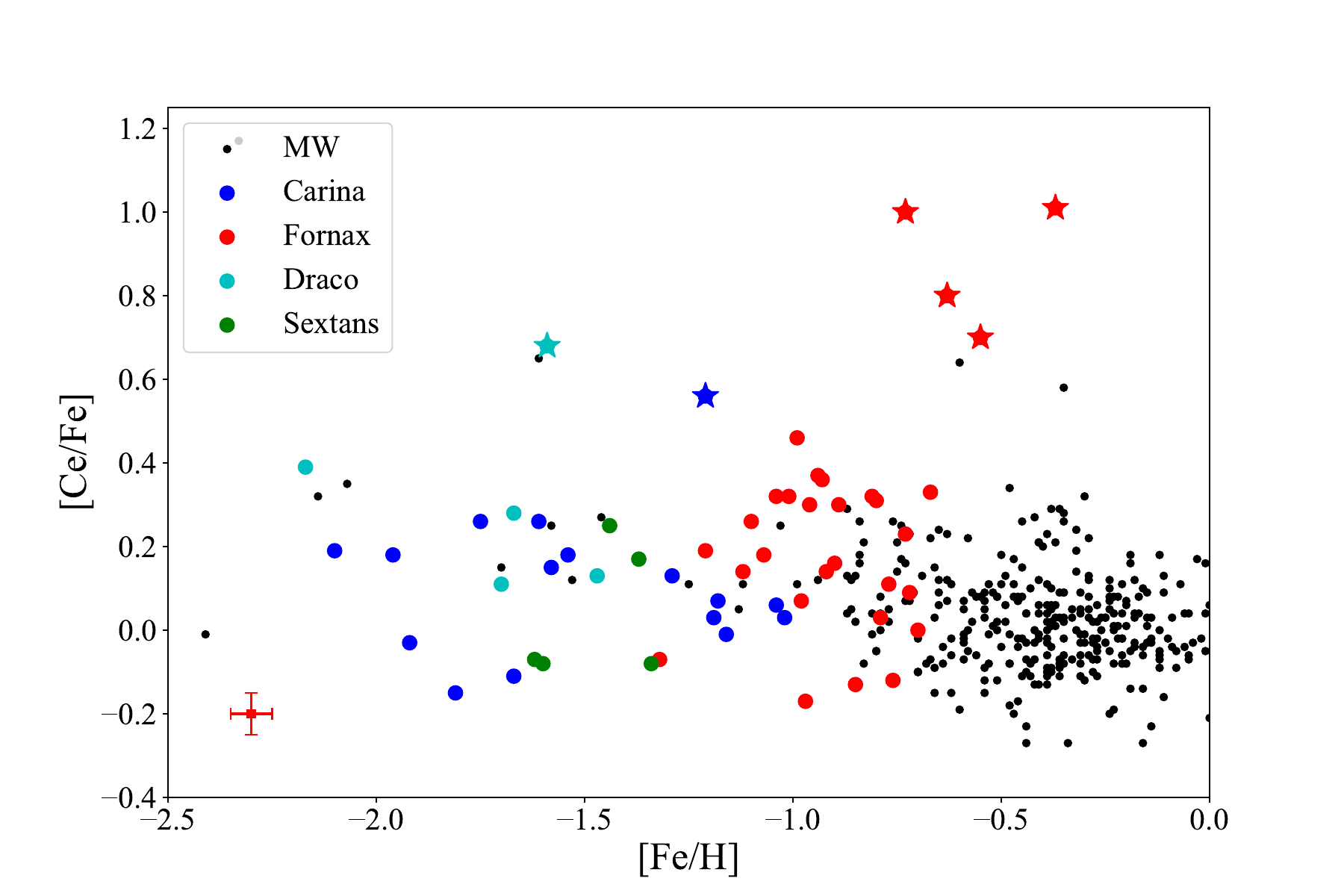} 
    \caption{[Ce/Fe] vs. [Fe/H]. Symbols are the same as in Fig.~\ref{fig: alpha}. Notably, the stars marked with star symbols are those with [Ce/Fe] greater than 0.5, which are likely AGB stars or have had AGB companions. }
    \label{fig: Ce-Fe}
\end{figure}

\subsection{Neutron-capture elements: Ce}
\label{sec: result-Neu}

\hspace{0.3cm}Elements heavier than iron are generally formed through neutron-capture processes, which are classified into rapid ({\it r}-) and slow ({\it s}-) neutron-capture processes depending on the relative rates of neutron capture and $\beta$-decay. For several decades, the astrophysical sites of these processes have been the subject of ongoing research and debate. For the {\it r}-process, current mainstream models include core-collapse SNe \citep[e.g.,][]{1994ApJ...433..229W} and neutron star mergers \citep[e.g.,][]{2018ApJ...855...99C,2019Natur.574..497W}. The s-process, on the other hand, primarily occurs in AGB stars during thermal pulsation phases \citep{2001ApJ...557..802B,2014PASA...31...30K}. 
According to \citet{2014ApJ...787...10B}, approximately 83\% of the Ce in our Sun originates from the {\it s}-process, which is similar to another frequently measured {\it s}-process-dominated element, Ba (85$\%$). 

\hspace{0.2cm}Six stars show exceptionally large [Ce/Fe] ($> +0.5$), including four from Fnx, one from Car, and one from Dra (Fig. \ref{fig: Ce-Fe}). One example of such Ce-rich star is shown in Fig. \ref{fig:lines}. The high [Ce/Fe] indicate that they are possible AGB stars or have had AGB companions. This is further confirmed by their higher values of [N/Fe] and [O/Fe] when available (see \citetalias{2023A&A...669A.125T} for further discussion). Interestingly, these Ce-rich stars are among the more metal-rich stars of their parent galaxies, particularly four Ce-rich Fnx stars show [Fe/H$] > -0.8$. The exact reason behind this is beyond the scope of this paper, but we speculate that it may be related to Fnx's high GC specific frequency (Section \ref{sec: discussion-nrich}), and thus frequent stellar interactions. After excluding the AGB candidates, we do not find significant [Ce/Fe] trend as a function of metallicity for all four dwarf galaxies, partially due to the large scatter in measured abundances. Notably, Ba, an element with similar {\it s}-process contribution as Ce, are measured in these four dwarf galaxies by \citetalias{2020A&A...641A.127R}. [Ba/Fe] generally show a flat trend as a function of metallicity for these dwarf galaxies between $-2.3<[$Fe/H$]<-0.5$, which is consistent with the null [Ce/Fe] trend in this work.

\section{Discussion}
\label{sec:discussion}

\subsection{Literature comparison and NLTE effects}

\begin{figure*}[htbp]
    \centering
    \includegraphics[width=0.85\textwidth]{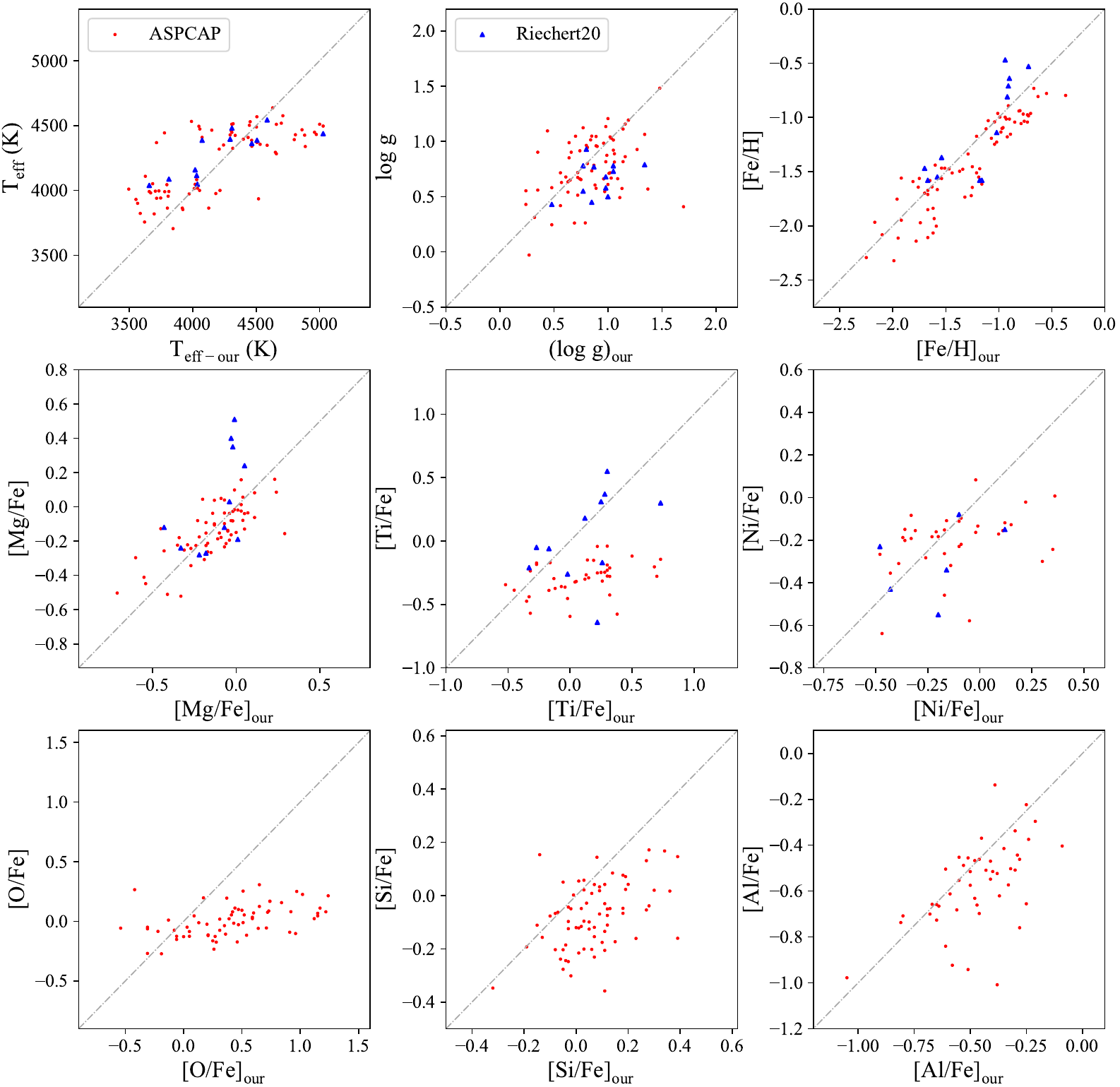}
    \caption{Comparison of stellar parameters and chemical abundances of our stars overlapping with those from \citetalias{2020A&A...641A.127R} (blue triangles) and APOGEE DR17 (red dots) respectively. Our results are denoted “our” and shown in the x-axis. The black dashed line is the 1:1 relation. }
    \label{fig: compare_new}
\end{figure*}

\hspace{0.3cm}In this section, we aim to verify our derived abundances and look for possible systematics by comparing with abundances derived from other counterparts. Here we pick the work of \citetalias{2020A&A...641A.127R} (we have 12 common stars with their work), which derived abundances of 380 stars in 13 dwarf galaxies with archival optical high-resolution spectra.  We also compare our derived abundances with the ASPCAP results from \cite{2021ApJ...923..172H}. 
The top panels of Fig. \ref{fig: compare_new} show that the stellar parameters are generally consistent across the three studies, albeit with non-negligible scatters. The log $g$ values from \citetalias{2020A&A...641A.127R} are systematically slightly smaller (by $\sim 0.3$ dex) than those from the other two studies. These offsets and scatters are expected, given the different methods used to determine stellar parameters (photometric versus spectroscopic; NIR versus optical).  
The middle and bottom panels compare the chemical abundances. The Ti and Ni, (as well as Cr and Mn\footnote{Not shown here due to limited space.}) abundances between our work and \citetalias{2020A&A...641A.127R} are consistent but exhibit significant scatters ($\sim 0.3-0.4$ dex). Compared to ASPCAP, our results show larger dynamical ranges for [Ti/Fe], [Si/Fe] and [O/Fe]. This likely reflects the different analysis techniques: our line-by-line analysis is more sensitive to variations in line strength than ASPCAP's spectral-window-based fitting.

\hspace{0.2cm}Our Mg abundances are systematically lower (approximately 0.2 dex) compared to those in \citetalias{2020A&A...641A.127R}, but similar to the ASPCAP results.  A similar offset is also found in \citetalias{2023A&A...669A.125T}. \citet{2021A&A...647A..24M} suggested 1D non-local thermodynamic equilibrium (NLTE) corrections for our adopted NIR Mg lines ($\lambda \sim 15741, 15749, 15766$ \AA) at the resolution of APOGEE spectra are about 0.2-0.4 dex, depending on different line and stellar parameters. Such NLTE corrections are expected to decrease the offset between our work and optical-derived abundances.

\hspace{0.2cm}The NLTE corrections for aluminum abundances measured by APOGEE are generally negative, ranging from about –0.1 to –0.2 dex for RGB stars. These corrections vary depending on the specific spectral lines used and the stellar parameters, but they do not significantly alter the separation between different stellar populations \citep{nlte_al_2023ApJ...953..143F}. Thus, while individual stellar abundances may be affected, the overall chemical trends found in this work remain robust. 

\hspace{0.2cm}The Mn I lines in the H band originate from high-excitation levels and are relatively sensitive to departures from LTE especially in the metal-poor giants \citep{nlte_mn_2008A&A...492..823B,nlte2020A&A...642A..62A}. The primary direction of the NLTE correction is positive. For metal-poor stars with [Fe/H] < -1.0, these corrections can reach between +0.2 dex and +0.5 dex (in extreme cases). Applying these NLTE corrections may raise the [Mn/Fe] ratios at low metallicities, possibly resulting in a flatter and potentially plateau-like trend in the [Mn/Fe] vs. [Fe/H] plane. For more discussions on NLTE effects of other elements, the readers are referred to dedicated works, e.g., \citet{NLTE_2020A&A...637A..80O,2021A&A...647A..24M,nlte_multi_2025A&A...699A..32K}. 

\begin{figure}[htbp]
    \raggedright
    \includegraphics[width=9.1cm,height=5.7cm]{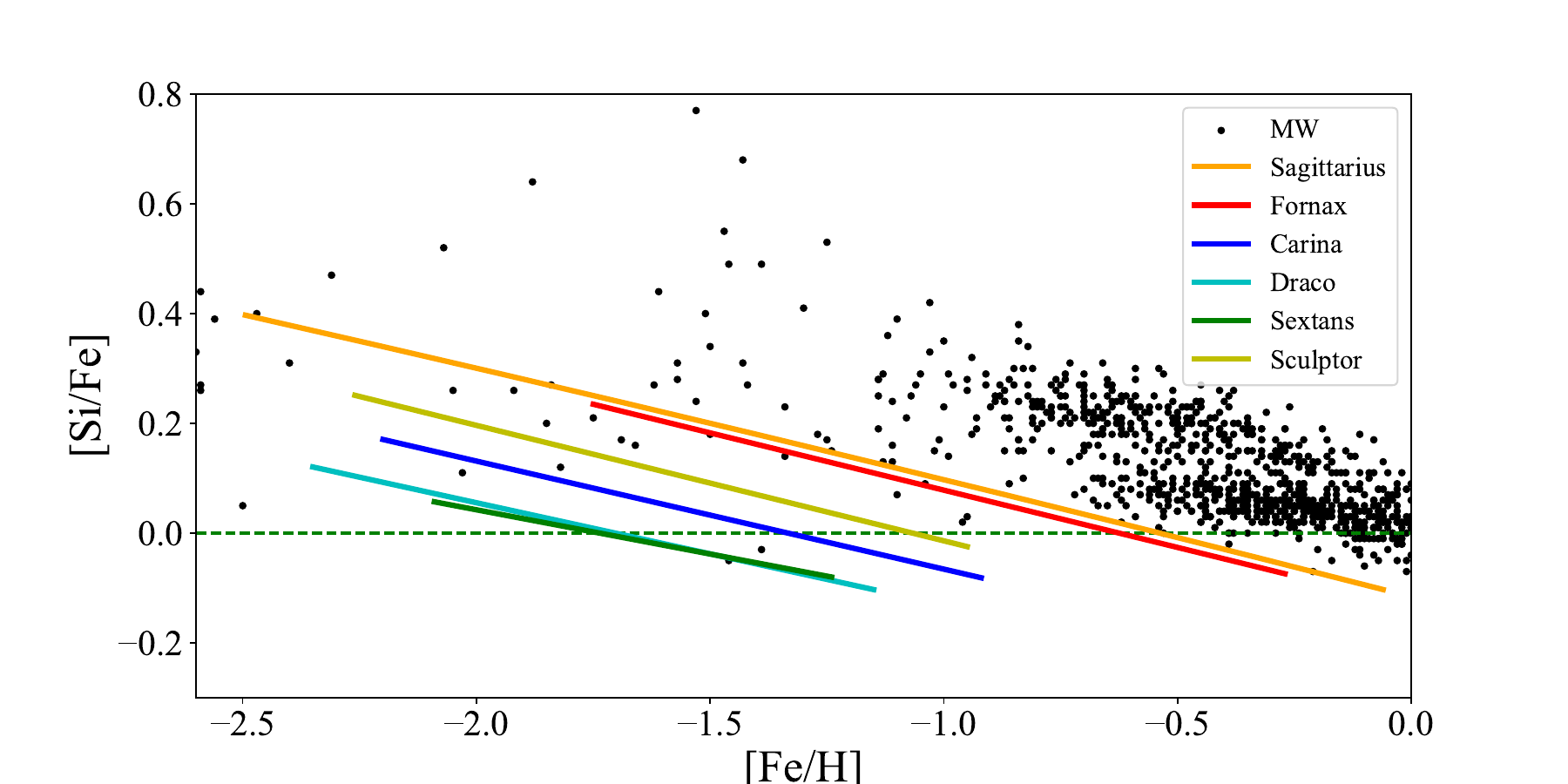} 
    \caption{Derived LTE abundances of [Si/Fe] over metallicities for individual dwarf galaxies. MW stars (black dots) are described in Fig. \ref{fig: alpha}. The best linear fits to the data of Sgr, Fnx, Dra, Car, Sex and Scl member stars are labeled as orange, red, cyan, blue, green, and yellow lines.}
    \label{fig: Si_trend}%
\end{figure}

\subsection{Galaxy mass and $\alpha$-element}
\label{sec:alphaknee}

\hspace{0.3cm}In general, more massive galaxies tend to be more metal-rich, a trend attributed to rapid and efficient star formation enrichment \citep[e.g.,][]{1979ApJ...229.1046T,2009ARA&A..47..371T}. It primarily reflects the extended SFHs of massive galaxies. Their deeper gravitational potential wells enable them to retain gas longer, leading to the prolonged formation of metal-rich stars with low [$\alpha$/Fe] ratios, as observed in the most massive Milky Way satellites. In this work, we collect the data of MW, Sgr, Scl, Fnx, Car, Dra, and Sex to find the rough correlation between the trend of $\alpha$-element and luminosity which is a proxy for galaxy mass. Given that O, Ca, Ti show larger scatters, and Mg may be systematically lower than those from the optical, we use Si as the tracer of $\alpha$-element (Fig. \ref{fig: Si_trend}). Limited by the sample sizes and metallicity coverage of each galaxy, accurate estimation of the $\alpha$-knee, if any, could be tricky. Instead, we perform a linear fit to the distribution of [Si/Fe] vs. [Fe/H] for each dwarf galaxy (Fig. \ref{fig: Si_trend}), and find out their intersection with [Si/Fe$]=0$ (Table \ref{Table:Si_trend}). Clearly, the more luminous, and thus more mass galaxies (Table \ref{Table:Si_trend}) show more metal-rich metallicities at the intersection, underscoring the critical role of galaxy mass in shaping chemical evolution. However, galaxy mass is not the only factor. Galaxy evolution is governed by a complex interplay of mass, environment, time, and internal/external processes. For example, Sagittarius and Fornax galaxies show similar metallicities at the intersection. However, given the substantial mass loss inferred for Sgr \citep{Majewski2003} and the absence of prominent tidal tails for Fnx \citep{2024A&A...690A..61D}, their similar luminosities (present-day stellar masses) should be interpreted with caution when inferring initial masses. While other mechanisms may also contribute to their similar $\alpha$ trends \citep{2009ARA&A..47..371T}, the luminosities nonetheless provide a useful, albeit imperfect, proxy for their initial mass scales.

\begin{table}

\caption{The metallicities at [Si/Fe] = 0 and absolute magnitudes for dwarf galaxies.}           
\label{Table:Si_trend}      
\centering                                      
\setlength{\tabcolsep}{5pt} 
\begin{tabular}{ccc}         
\hline\hline
\\[-3pt]

Galaxy & [Fe/H]$_{[Si/Fe]=0}$ & M$_V$ [mag]$^a$ \\

\hline\\ 
Sgr & $-0.55 \pm \ 0.16$ & $-13.5 \pm \ 0.30$\\
Fnx & $-0.62 \pm \ 0.15$ & $-13.46 \pm \ 0.14$\\
Scl & $-1.08 \pm \ 0.11$ & $-10.82 \pm \ 0.14$\\
Car & $-1.35 \pm \ 0.23$ & $-9.43 \pm \ 0.05$\\
Sex & $-1.71 \pm \ 0.27$ & $-8.72 \pm \ 0.06$\\
Dra & $-1.72 \pm \ 0.25$ & $-8.71 \pm \ 0.05$\\

\hline                                   
\end{tabular}
\tablefoot{ $^a$: From \citet{2012AJ....144....4M} and \citet{2018ApJ...860...66M}.}

\end{table}

\subsection{Radial chemical gradients}
\label{sec: discussion-mg}

\begin{table}
        \caption{Results of linear least squares regression for Fnx}
        \label{tab:XFe_a_linearFitting_coefficient}
        \begin{center}
\begin{tabular}{c c c c}
\hline
\hline
\hline
element & k & k error & p-value \\
    & (dex/degree) &(dex/degree) &  \\
\hline
\hline
  {\rm [Fe/H]}  & $-0.7814$ & 0.217 & 0.0004\\
  {\rm [O/Fe]}  & $-0.2480$ & 0.488 & 0.615\\
  {\rm [Mg/Fe]} & $-0.1475$ & 0.221 & 0.510\\
  {\rm [Al/Fe]} & $-0.2178$ & 0.229 & 0.351\\
  {\rm [Si/Fe]} & 0.0687  & 0.149 & 0.647\\
  {\rm [Ca/Fe]} & $-0.1202$ & 0.215 & 0.581\\
  {\rm [Ti/Fe]} & 0.2733  & 0.390 & 0.489\\
\hline
\hline
\end{tabular}
        \end{center}
        \tablefoot{ k: best-fit slope. p-value: two-sided p-value for a hypothesis test in which the null hypothesis is that the slope is zero.}
\end{table}

\hspace{0.3cm}Radial chemical gradients are important observational constraints to reconstruct galaxy formation history \citep{2017A&A...603A...2M}.  The sample sizes of Dra, Car, and Sex are not large enough for meaningful discussions, and we focus on Fnx in this section. Being one of the first dwarf spheroidals discovered by Harlow Shapley in 1938, Fnx has been investigated for a long time. Fnx shows a prolonged SFH, with at least three distinct stellar populations: a young population (few 100 Myr old), an intermediate age population (2-8 Gyr old), and an ancient population ($>10$ Gyr, \citealt{2006A&A...459..423B}). The stellar populations in Fnx can be separated into two/three components considering their spatial, kinematics and metallicity distributions: the metal-rich component is more centrally concentrated with colder kinematics, and vice versa \citep{2006A&A...459..423B,2011ApJ...742...20W,2012ApJ...756L...2A}. These components show misalignment between their angular momentum vectors and different anisotropy parameters \citep{2012ApJ...756L...2A, 2022A&A...659A.119K}, indicating a  ``core'' dark matter density profile \citep{2011ApJ...742...20W}, and a late merger history \citep{2012ApJ...756L...2A}.

\begin{figure*}
        \sidecaption
        \includegraphics[width=12cm,angle=0]{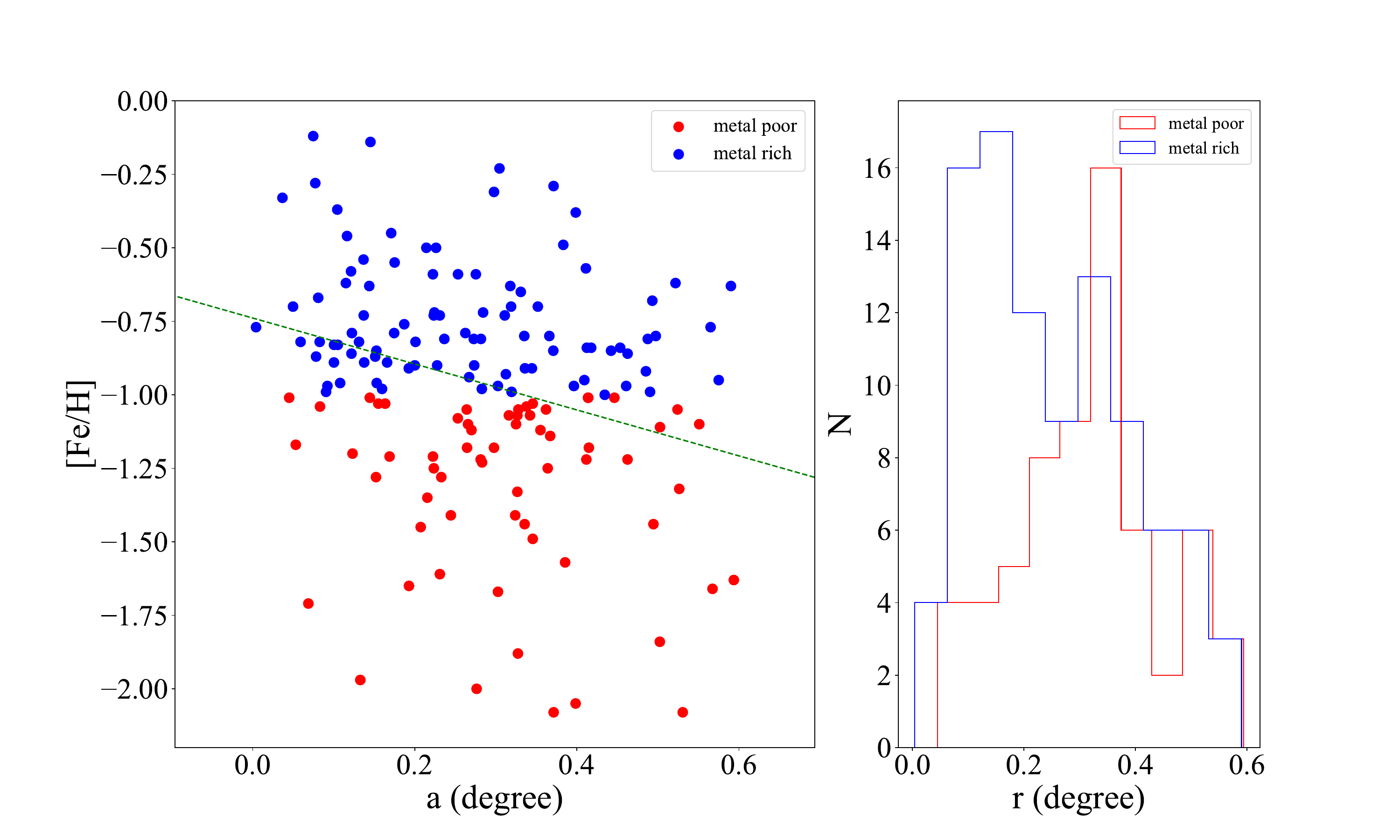}
        \caption{Metallicity distribution of stars in Fnx. Left panel: {\rm [Fe/H]} as a function of elliptical radius in degrees. The green dashed line represents the best least-squared fit. They are divided into a metal-rich population ([Fe/H$]>-1$, blue dots) and a metal-poor population ([Fe/H$]<-1$, red dots).
        Right panel: the radial distribution of the metal-rich population (blue histogram) and the metal-poor population (red histogram).
        }
        \label{fig:metal_bin}
\end{figure*}

Chemical abundance investigation is more challenging, since it demands spectra with higher SNR (\citealt{2003AJ....125..684S}; \citetalias{2020A&A...641A.127R}). \citet{2021ApJ...923..172H} suggests a second, weaker ``burst'' star formation in Fnx based on chemical modelling the [$\alpha$/Fe]-[Fe/H] patterns, in agreement with results from the color-magnitude diagram \citep{boer_2012A&A...539A.103D}. As chemical abundances are strongly affected by galaxy SFH, here we explore the possible radial chemical gradient in Fnx for the first time. 
We used linear least-squares regression fitting to find the correlations between chemical abundances and elliptical galactic radius. We list the results in Table \ref{tab:XFe_a_linearFitting_coefficient}, including their slopes (k), slope errors, and p-value\footnote{A hypothesis test is performed to assess whether the slope equals zero, using the Wald Test with a t-distribution for the test statistic.}. Our results suggest a nondetectable radial gradient for [O/Fe], [Mg/Fe], [Al/Fe], [Si/Fe], [Ca/Fe] and [Ti/Fe], given their two-sided p-values are too large ($>0.05$). However, we confirmed the metallicity gradient (p-value$= 0.0004$ and $k=-0.7814$, left panel of Fig. \ref{fig:metal_bin}) suggested in literature \citep{2006A&A...459..423B,2012ApJ...756L...2A}, where we further include stars with $40<$ SNR $<70$.  If we divide them with [Fe/H$]=-1$ according to \citet{2022A&A...659A.119K}, we find the average radius of the metal-rich stars is 0.1-0.2 degree, while it is 0.3-0.4 degree for their counterparts (Right panel of Fig. \ref{fig:metal_bin}). Thus the metal-rich stars are statistically located closer to the galactic center. This agrees with \citet[][Fig. 15]{2006A&A...459..423B}.
The lack of detectable radial gradients of element abundance ratios could be attributed to the following reasons: 1. Not large enough sample size (32 Fnx stars); 2. Measurement uncertainties and offsets ($\sim 0.1-0.2$ dex); 3. Three star-formation epochs and possible merger history may form more complicated radial chemical gradients. When a new large-scale SF event occurs, one would expect the [$\alpha$/Fe] to increase after the first new SNe II and for the first 0.1-1 Gyrs until SNe Ia kick in, complicating any abundance gradient using $\alpha$-elements. We speculate the last one may be the driving factor: radial gradients of [Mg/Fe] and [Ca/Fe] were identified in Scl \citepalias{2023A&A...669A.125T}, which formed most of the stars early with a single peak \citep{monki_1999PASP..111.1392M,boer_2012A&A...539A.103D,betti_2019MNRAS.487.5862B,2022ApJ...925...66D}.

\subsection{N-rich field stars}
\label{sec: discussion-nrich}

\hspace{0.3cm}As part of our on-going project ``{\bf GASTRONOMI}'', we aim to investigate the connection between star cluster properties and galaxy chemical evolution in these lower mass MW satellites. A substantial number of the existing MW GCs were born ex-situ, and later joined other in-situ GCs by mergers \citep[e.g.,][]{massari2019}. GCs are also found in large MW satellite galaxies (e.g., LMC and SMC) and smaller dwarf galaxies \citep{2021MNRAS.500..986H}. Many GCs are dissolved after birth \citep{2019MNRAS.486.4030L}, while member stars of surviving GC may escape its potential well through two-body relaxation. Therefore, GC escapees are expected to be found in the field. Using the chemical anomaly found in GCs, astronomers identified N-rich field stars as possible candidates for GC escapees. These stars can be found in dissolved dwarf galaxies (e.g., GSE, \citealt{2021ApJ...913...23Y}), dissolving dwarf galaxies (e.g., Sgr, \citealt{2021A&A...648A..70F}), and existing dwarf galaxies (e.g, LMC \& SMC, \citealt{2020ApJ...903L..17F}).

\begin{figure}
        \raggedright 
        \includegraphics[width=9.05cm,height=6.85cm]{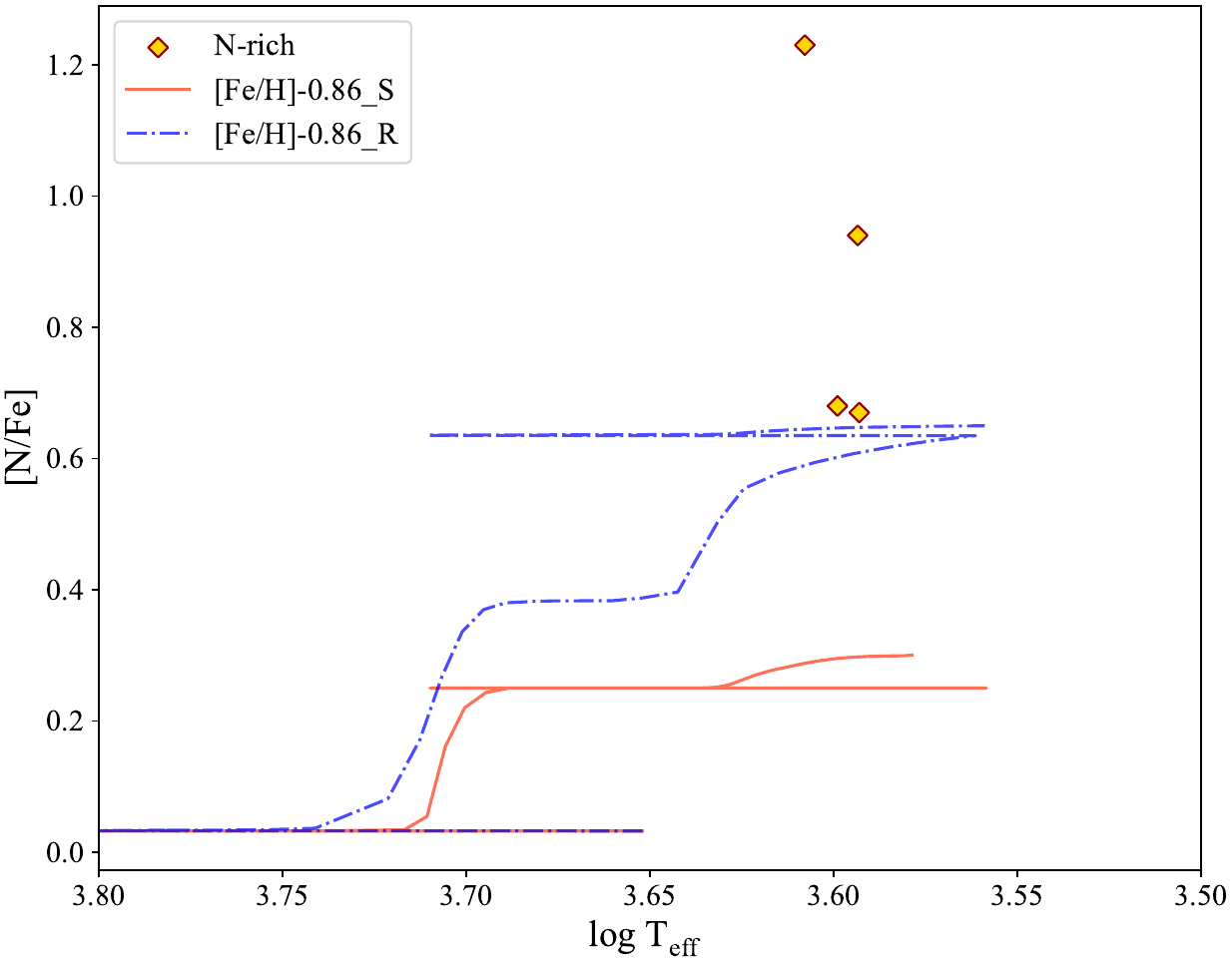}
        \caption{Comparison between our identified N-rich stars (gold diamonds) and model predictions of \cite{lagarde_2012A&A...543A.108L}. Red and blue lines represent predictions for the stellar mass of 1.0 $\mathrm{M_{\odot}}$ in the standard model and the model with thermohaline convection and rotation-induced mixing respectively (Z = 0.02, [Fe/H] = $-0.86$). }
        \label{fig:model}
\end{figure}

\begin{figure}
        \raggedright \includegraphics[width=8.8cm,height=6.6cm]{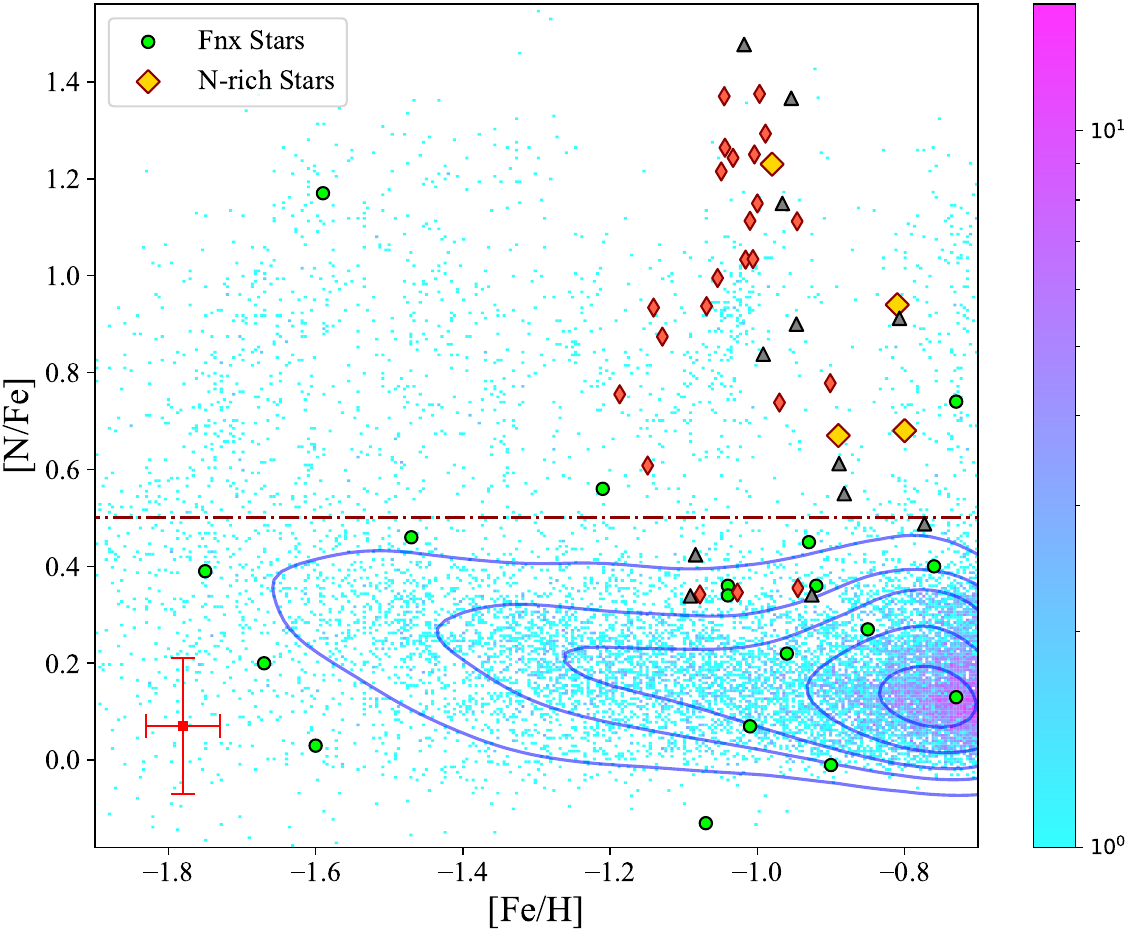}
        \includegraphics[width=8.8cm,height=6.6cm]{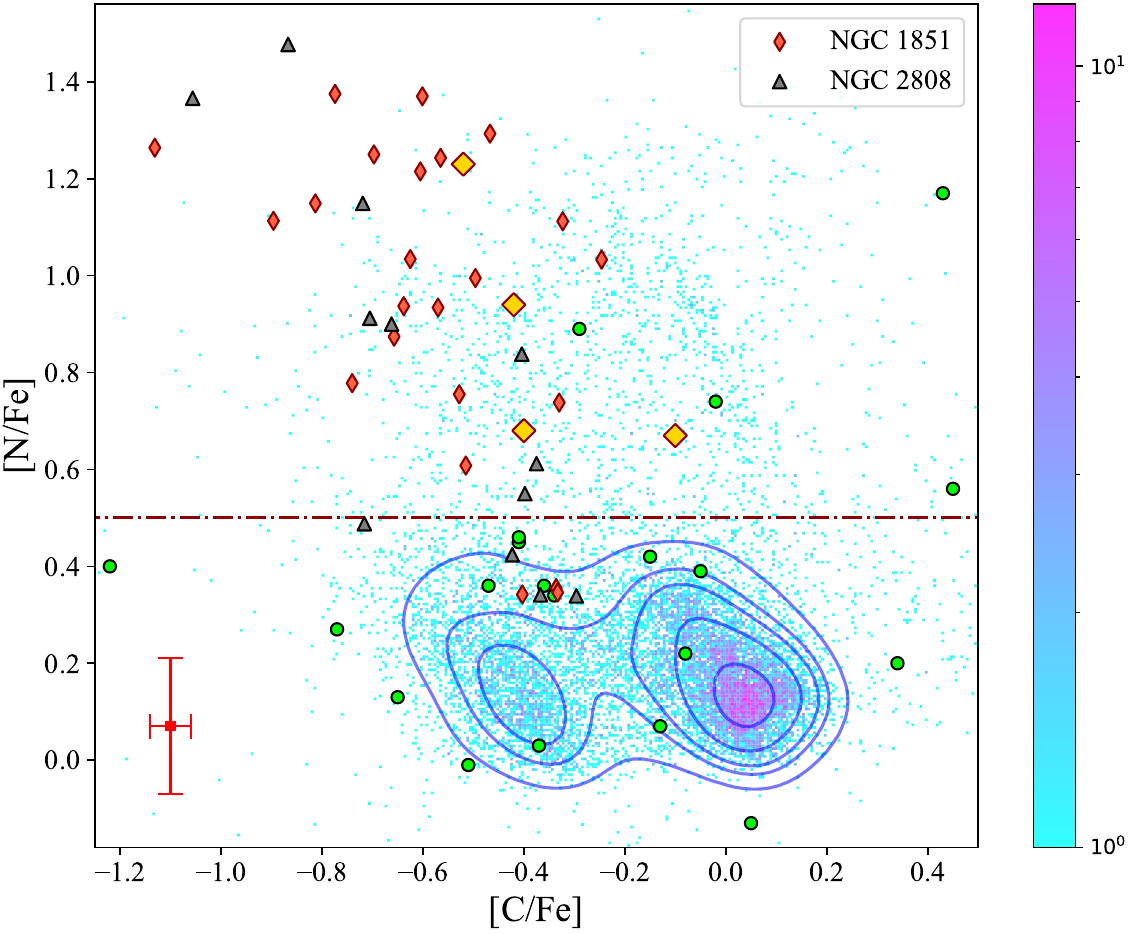}
        \caption{[N/Fe] vs. [Fe/H] and [N/Fe] vs. [C/Fe] planes. Stars from the Fnx dwarf galaxy is shown as lime circles, while our identified N-rich stars are labeled as golden diamonds.  Light red diamonds stand for stars in NGC 1851 and gray triangles for NGC 2808 \citep{Meszaros2020}. Colored dots and blue contours correspond to metal-poor ([Fe/H] < $-0.7$) MW RGB stars from APOGEE DR17. }
        \label{fig:nfe}
\end{figure}

Next, we try to identify N-rich field stars in Fnx, Dra, Car, Sex dwarf galaxies with the following criteria: [N/Fe$]>0.65$, [C/Fe$]<0.15$, and [Ce/Fe$]<0.5$. Here we exclude stars with high C and Ce abundances to avoid AGB star contamination \citep{2017MNRAS.465..501S, 2019MNRAS.488.2864F}. We identify 4 N-rich field stars out of 32 sample stars (4/32) in Fnx, 0/14 in Dra, 0/19 in Car, and 0/8 in Sex. Considering that the effects of internal mixing in giant stars could also cause high N surface abundances, we compared the [N/Fe] and $\mathrm{T_{eff}}$ of these N-rich stars with the stellar models mentioned in \cite{lagarde_2012A&A...543A.108L} (see Fig.\ref{fig:model}). Models including thermohaline convection and rotation-induced mixing are considered here. The effects of internal mixing on surface [N/Fe] are predicted to become more pronounced at lower metallicities. This explains the high [N/Fe] ratios observed by \cite{spite_2005A&A...430..655S} in their sample of extremely metal-poor halo stars ([Fe/H] $\lesssim$ -3). In contrast, the N-rich stars identified in our work reside at significantly higher metallicities ([Fe/H] > -1), where the internal mixing effects alone are insufficient to account for the observed N enhancements. Instead, such stars are more likely to be formed from gas previously enriched by the CNO-processed ejecta of massive first-generation stars, imprinting higher [N/Fe] at birth.  This statement is further supported by comparing these N-rich field stars with MW metal-poor giants and GC stars (Fig. \ref{fig:nfe}). The N-rich field stars show much higher [N/Fe] compared to the bulk MW RGB stars\footnote{Also note that our N-rich field stars have lower log $g$ than the majority of MW RGB stars.}, but resemble those N-rich stars from GCs with similar metallicities (NGC 1851 and NGC 2808). 

Additionally, N-rich field stars may show abnormal abundances in other elements, e.g., Al \citep{2021ApJ...913...23Y, 2021A&A...648A..70F}. However, we do not find such peculiarity in the four N-rich field stars of Fnx. Specifically, none of them show Al enrichment, which does not contradict the hypothesis that they are escaped from GCs ----  more metal-rich GCs usually do not show Al enrichment \citep{Pancino2017}. 
The lack of N-rich field stars in Dra, Car, and Sex is not surprising, as similar situation is found in Scl \citepalias{2023A&A...669A.125T}. It is probable that the low-mass, diffuse galactic environment is not suitable for the formation of GCs, and therefore should not have any N-rich field stars if they were formed in GCs.

Fnx shows not only a special chemodynamical structure (Section \ref{sec: discussion-mg}), but also remarkably high GC specific frequency\footnote{The number of GCs per unit galaxy luminosity, i.e., $S_N= N_{GC} \times 10^{(M_V+15)}$.}, $S_N=26$. This value is exceptionally high among local galaxies, but it is comparable to those found in dwarf galaxies of the Coma cluster \citep{2020MNRAS.492.4874F}. 
Currently, six GCs are suggested to be associated with Fnx (e.g., see \citealt{2019ApJ...875L..13W,2021ApJ...923...77P} and references therein), where three of them, Fnx 1 ([Fe/H$]=-2.5 \pm 0.1$), Fnx 2 ([Fe/H$]=-2.1 \pm 0.1$), Fnx 3 ([Fe/H$]=-2.4 \pm 0.1$), show MPs, i.e., star-to-star variations in O, Na, and Mg \citep{2006A&A...453..547L}. Moreover, Fnx 3 ([Fe/H$]=-2.28 \pm 0.02$), Fnx 4 ([Fe/H$]=-1.24 \pm 0.02$), and Fnx 5 ([Fe/H$]=-2.06 \pm 0.03$) were investigated by integrated-light spectra \citep{2022A&A...660A..88L}. Fnx 6 is the most metal-rich one with [Fe/H$]=-0.71 \pm 0.05$ \citep{2021ApJ...923...77P}.

Given the metallicity range of our measured Fnx stars, we plot the abundances of Fnx 3, Fnx 4, and Fnx 5 \citep{2022A&A...660A..88L} in Fig. \ref{fig: alpha} and \ref{fig: Iron-Fe} to visualize the abundance trend at lower metallicities. With similar metallicity, Fnx 4 perfectly matches with our Fnx field stars in all measured abundances. Fnx 3 and Fnx 5 also match with other Fnx field stars in elements heavier than Ca, but lower [Mg/Fe] and higher [Si/Fe] are spotted when compared with field stars of other galaxies (e.g., MW). These peculiar features are in fact related to their GC nature: 1. The star-to-star Mg variation with depleted values is found in Fnx 3 \citep{2006A&A...453..547L}, where a similar situation may also be true in Fnx 5; 2. Enhanced Si abundances are commonly found in metal-poor Galactic GCs \citep[e.g.,][]{2018ApJ...855...38T}. Therefore, the enhanced Si integrated-light abundances indicate the existence of star-to-star Si variation with elevated values in Fnx 3 and Fnx 5. Our identified N-rich field stars show metallicity between $-0.8$ and $-1.0$ (Table B2) with [Fe/H] uncertainties of $\sim 0.05$ dex. They do not overlap with any existing Fnx GCs either in metallicity or in spatial location (Fig. \ref{fig:FornaxCMD}). 
Assuming that N-rich field stars are escaped from GCs \citep{2020ApJ...891...28T,2021ApJ...913...23Y}, this may indicate GC destruction in Fnx. 
In fact, \citet{2023MNRAS.522.5638C} predicted the number of GCs drops by a factor of $\sim 5-8$ compared to the peak value based on modeling massive satellites with similar properties as Fnx. Therefore, the existence of GC escapees is expected by theoretical models. To further verify the destruction rate, a complete census of Fnx stars with high accurate abundances is needed.

\section{Summary}
\label{sec: summary}

\hspace{0.3cm}Although dwarf galaxies outnumber bright galaxies, their low surface brightness makes them challenging to observe in the distant Universe. The satellite dwarf galaxies orbiting the MW serve as ideal laboratories for studying the detailed evolution of galaxy systems. Specifically, we can model their evolution using resolved stellar populations in color-magnitude diagrams and their chemical abundances. Additionally, we can investigate the coevolution of galaxies and star clusters, as these systems mutually influence one another. Following our initial study of the Scl dwarf galaxy, this paper represents our second investigation into low-mass MW satellites using high-resolution spectra. 

We derived chemical abundances for Fe, C, N, O, Mg, Al, Si, Ca, Ti, Cr, Mn, Ni, and Ce in 74 member stars across the Fnx, Dra, Car, and Sex dwarf galaxies. To address the weak spectral lines in these metal-poor stars, we determined abundances line-by-line using the Turbospectrum wrapper --- Bacchus.
\begin{enumerate}
\item The observed abundances of $\alpha$ elements (O, Mg, Si, Ca, Ti) are consistent with the ``flat then decreasing'' trend with increasing metallicity. Given the smaller uncertainties and negligible systemic offset display by Si, the distribution of [Si/Fe] shows a strong correlation with the absolute magnitudes of galaxies (including Sgr, Scl and four dwarf galaxies in this work), confirming the critical role of galaxy mass in shaping chemical evolution.  
\item Al abundances in the studied dwarf galaxies cluster around [Al/Fe]$\sim -0.5$, consistent with values observed in the Sgr dwarf galaxy or MW metal-poor stars. While Car, Dra, and Sex show no discernible metallicity-dependent trends, Fornax displays a tentative flattening of [Al/Fe] near [Fe/H]$\sim-0.75$ after an initial decline with increasing metallicity.
\item Leveraging Fnx’s larger stellar sample, we investigated radial chemical abundance gradients. While confirming the previously reported metallicity gradient, we detected no significant radial trends for O, Mg, Al, Si, Ca, or Ti. This may reflect Fnx's complex star-formation history and potential merger events.
\item N-rich field stars --- potential escapees from globular clusters (GCs) formed in dense environments --- are absent in Dra, Car or Sex, consistent with their lack of GC as low-mass dwarfs. However, N-rich field stars were identified in Fnx. These stars exhibit no anomalous abundances in other measured elements and show metallicities distinct from known Fnx GCs, suggesting they may represent remnants of one or more dissolved GCs. 
\end{enumerate}
In the coming years, next-generation astronomical facilities --- such as the Extremely Large Telescope (ELT) and multi-fiber spectroscopic surveys (e.g., 4MOST, WEAVE, DESI) --- will deliver increasingly detailed chemical profiles of member stars within dwarf galaxies. These advancements will enable tighter constraints on abundance patterns at the metal-poor end, a critical input for modeling the early evolutionary stages of these systems. By conducting comprehensive studies of dwarf galaxies across a range of mass scales, we will gain transformative insights into their formation histories and dynamical interactions, significantly advancing our understanding of galaxy evolution and the cosmic processes shaping these faint stellar systems.

\section{Data availability}
Tables B1 and B2 are only available in electronic form at the CDS via anonymous ftp to cdsarc.u-strasbg.fr (130.79.128.5) or via http://cdsweb.u-strasbg.fr/cgi-bin/qcat?J/A+A/.

\begin{acknowledgements}
C.X., Y.Q and B.T. gratefully acknowledge support from the National Natural Science Foundation of China through grants NOs. 12233013 and 12473035, China Manned Space Project under grant NO. CMS-CSST-2025-A13 and CMS-CSST-2021-A08, the Fundamental Research Funds for the Central Universities, Sun Yat-sen University (24qnpy121). J.G.F-T gratefully acknowledges the support provided by ANID Fondecyt Regular No. 1260371, ANID Fondecyt Postdoc No. 3230001 (Sponsoring researcher), the Joint Committee ESO-Government of Chile under the agreement 2023 ORP 062/2023 and the support of the Doctoral Program in Artificial Intelligence, DISC-UCN. Z.Y. acknowledges the support from the National Natural Science Foundation of China under grant numbers 12203021.
D.G. gratefully acknowledges the support provided by Fondecyt regular no. 1220264.
D.G. also acknowledges financial support from the Direcci\'on de Investigaci\'on y Desarrollo de
la Universidad de La Serena through the Programa de Incentivo a la Investigaci\'on de
Acad\'emicos (PIA-DIDULS).

\end{acknowledgements}

%
%

\bibliographystyle{aa}
\bibliography{MW_satellite.bib}


\begin{appendix} 
\appendix
\onecolumn
\centering

\section{Example Key Spectral Lines}

\begin{figure*}[htbp]
    \centering
    \includegraphics[width=18.5cm, height = 14.5cm]{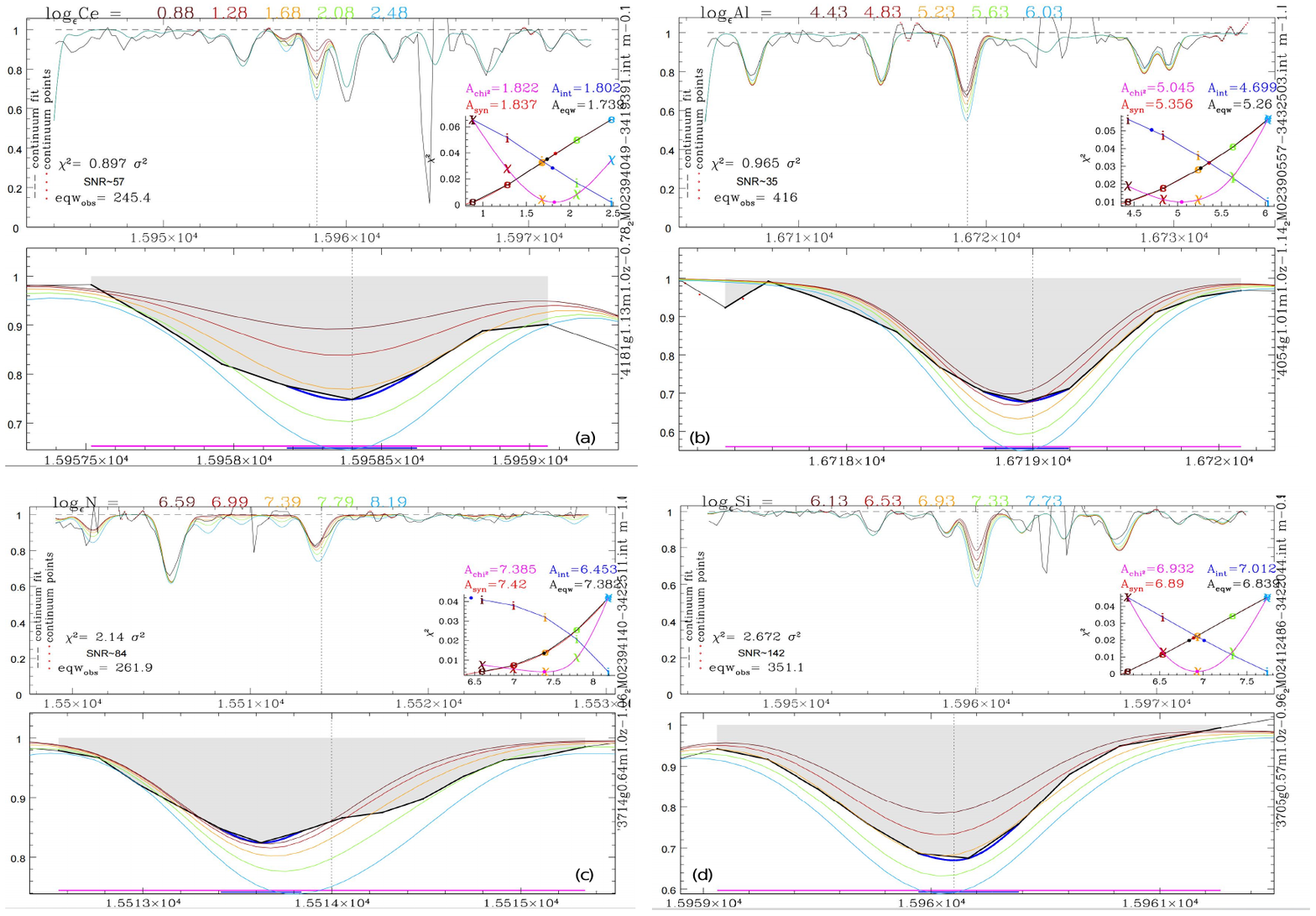}
    \caption{Key lines utilized to calculate chemical abundances in BACCHUS (take Ce, Al, N, Si as example for neutron-capture elements, odd-Z elements, light elements and $\alpha$ elements). }
    \label{fig:lines}
\end{figure*}

\section{Information on all sample stars}
\label{app:sample}
Basic information on all sample stars (Table B1) and their chemical abundances (Table B2) are available at the CDS.

\end{appendix}
\end{document}